\documentclass{ws-procs9x6}

\def\beq{\begin{equation}}
\def\eeq{\end{equation}}
\def\br{\begin{eqnarray}}
\def\er{\end{eqnarray}}
\def\benu{\begin{enumerate}}
\def\eenu{\end{enumerate}}

\def\l{\left}
\def\r{\right}    
\def\Hbar{\mathcal H}

\begin{document}

\title{Planck scale effects and the suppression of power
on the large scales in the primordial spectrum}
\author{S.~Shankaranarayanan}
\address{HEP Group, ICTP, Strada costiera 11, 34100 Trieste, Italy.\\
E-mail: shanki@ictp.trieste.it}
\author{L.~Sriramkumar}
\address{Harish-Chandra Research Institute, Chhatnag Road,
Jhunsi, Allahabad, India.\\ 
E-mail: sriram@mri.ernet.in}

\maketitle

\abstracts{The enormous red-shifting of the modes during the inflationary 
epoch suggests that physics at the very high energy scales may modify
the primordial perturbation spectrum.  Therefore, the measurements of
the anisotropies in the Cosmic Microwave Background (CMB) could
provide us with clues to understanding physics beyond the Planck
scale.  In this {proceeding}, we study the Planck scale effects on
the primordial spectrum in the power-law inflation using a model which
preserves local Lorentz invariance.  While our model reproduces the
standard spectrum on small scales, it {\it naturally}\/ predicts a
suppression of power on the large scales---a feature that seems to be
necessary to explain deficit of power in the lower multipoles of the
CMB.}

\section{Introduction}

In the inflationary scenario\cite{Linde:1990-bk}, the perturbations 
corresponding to comoving length scales of cosmological interest 
today would have emerged from quantum fluctuations at the beginning 
of inflation with physical wavelengths smaller than the Planck length. 
Hence, in principle, quantum gravitational effects should have left 
their signatures on the primordial spectrum. 
This opens up the interesting possibility of probing trans-Planckian (TP) 
physics using the CMB~\cite{Brandenberger-Mart:2000}.

The first year results of WMAP\cite{Peiris-WMAP:2003} data show that
the power in the quadrupole and the octopole moments of the CMB are 
{\it lower than}\/ as expected in the best fitting $\Lambda CDM$ models.  
The deficit of power in the lower multipoles can not be explained within 
the context of the standard inflationary models (unless these models are
fine-tuned\cite{Contaldi-Pelo:2003}) and suggests a possible signature 
of TP physics.

Most of the earlier efforts in incorporating TP physics into the standard 
field theory have involved models which {\it break}\/ local Lorentz 
invariance\cite{Brandenberger-Mart:2000}. 
However, theoretically, there exists no apriori reason to believe that 
Lorentz invariance may be broken at the scales of inflation. 
More importantly, recent observations of 
synchrotron emission 
from the Crab nebula 
seem to suggest that Lorentz invariance may be preserved to very high 
energies\cite{Jacobson-Libe:2002}. 
In such a situation, in order to study the TP effects on the primordial 
perturbation spectrum, it becomes important that we also consider models 
which preserve Lorentz invariance {\it even as they contain a fundamental 
scale}.\/ 
In this proceeding, we consider one such model, evaluate the resulting
spectrum of density perturbations in power-law inflation and also discuss 
its implications for the CMB angular power spectrum.

\section{The model and its application to power-law inflation}

In the inflationary scenario, the primordial perturbation spectrum per
logarithmic interval, viz. $\left[k^3\, {\mathcal P}_{\Psi}(k)\right]$, 
is given by\cite{Linde:1990-bk} 
\beq
\int\limits_{0}^{\infty} d(\ln k) \, 
\left[k^3\; {\mathcal P}_{\Psi}(k)\right] 
= G^{+}_{0}\l({\tilde x},{\tilde x}\r),
\label{eq:psdfntn}
\eeq 
where $G^{+}_{0}({\tilde x}, {\tilde x'})$ denotes the Wightman function 
corresponding to a massless and minimally coupled, quantum scalar field 
(say, ${\hat \Psi}$) evolving in the inflating background and the spectrum
is to be evaluated at Hubble exit.  
Therefore, in order to understand the effects of Planck-scale physics 
on the perturbation spectrum, we need to understand as to how quantum 
gravitational effects will modify the propagator of a massless scalar 
field in the inflationary background.

Motivated by the Pauli-Villars regularization procedure, we assume that 
the massless Wightman function, viz. $G_{0}^{+}\l({\tilde x},{\tilde x'}
\r)$, is modified due to TP effects (in a locally Lorentz invariant manner) 
to\cite{Shanki-Srir:2004}
\beq
G_{\rm M}^{+}\l({\tilde x},{\tilde x'}\r) 
= G_{0}^{+}\l({\tilde x},{\tilde x'}\r) 
- G_{k_{\rm c}}^{+}\l({\tilde x},{\tilde x'}\r),\label{eq:mGfnx}
\eeq
%
%
%
%
where $G_{k_{\rm c}}^{+}\l({\tilde x},{\tilde x'}\r)$ is the Wightman 
function of a massive scalar field of mass $k_{\rm c}$ and $k_{\rm c}$ 
denotes the cut-off scale which we shall assume to be three to five orders 
of magnitude above the Hubble scale during inflation. 
Then, following Eq.~(\ref{eq:psdfntn}), we can define the resulting 
modified perturbation spectrum, viz.~$\left[k^3\; {\mathcal P}_{\Psi}(k)
\right]_{\rm M}$, as follows:
\beq 
\int\limits_{0}^{\infty} 
d(\ln k)\, \left[k^3\; {\mathcal P}_{\Psi}(k)\right]_{\rm M}
= G_{\rm M}^{+}\l({\tilde x},{\tilde x}\r)
= G_{0}^{+}\l({\tilde x},{\tilde x}\r) 
- G_{k_{\rm c}}^{+}\l({\tilde x},{\tilde x}\r).
\label{eq:mpsdfntn}
\eeq

Let us now consider the massless and massive scalar fields to be
propagating in a power-law inflationary background described by 
the line element
\beq
ds^2 =a^{2}(\eta)\l(d\eta^{2} - d{\bf x}^2\r),\label{eq:frw}
\eeq 
where $\eta$ is the conformal time, $a(\eta) = \l(-\Hbar \,
\eta \r)^{(\beta+1)}$ with $\beta \le -2 $ and $\Hbar$ denotes the 
energy scale associated with inflation.
If we assume that both the fields are in the Bunch-Davies vacuum, then 
the modified power spectrum~$\left[k^3\; {\mathcal P}_{\Psi}(k)
\right]_{\rm M}$ can be expressed as
\beq 
\left[k^3\; {\mathcal P}_{\Psi}(k)\right]_{\rm M} 
=\frac{k^3}{2\pi^2 \, a^2}\, 
\l(\vert \mu_{k}\vert^2  
- \vert{\bar \mu}_{k}\vert^2\r),\label{eq:psm}
\eeq 
where $\mu_{k}$ and ${\bar \mu}_{k}$ denote the normal modes of the 
massless and the massive fields satisfying the differential equations 
\br
\mu_{k}''+\l[k^2 - a''/a\r]\mu_{k}&=&0,\label{eq:demu0}\\
%
{\bar \mu}_{k}''+\l[k^2+(k_{\rm c}\, a)^2
- a''/a\r]{\bar \mu}_{k} &=&0,\label{eq:demukc}
\er
respectively.
On comparing Eqs.~(\ref{eq:psdfntn}) and (\ref{eq:psm}), it is clear
that, in our model, the TP corrections to the standard spectrum arise 
as a result of the contribution due to the massive modes.
Before we proceed further with the evaluation of the corrections, we 
need to stress the following point: In the standard inflationary scenario, 
it is well-known that the amplitude of the spectrum corresponding to the 
massive modes decays at super-Hubble scales. 
In our model, the TP corrections to the standard spectrum are due to the 
massive modes. 
Hence, within the standard inflationary picture, the amplitude of these
corrections would be expected to decay at super-Hubble scales. 
However, as the massive modes we have considered are supposed to represent 
TP corrections to the standard massless modes, in what follows, we shall 
assume that the mechanism that `freezes' the amplitude of the standard
spectrum at super-Hubble scales will also `freeze' the amplitude of the 
corrections at their value at Hubble exit.

The mode functions for the massless field in power-law inflation can be
expressed in terms of Hankel functions and the standard power-spectrum, 
evaluated at Hubble exit, is given by\cite{Linde:1990-bk}
%
\beq
\label{eq:psplfv}
\left[k^3\; {\mathcal P}_{\Psi}(k)\right] 
= C \, \l(\frac{\Hbar^2}{2 \pi ^2}\r) \, 
\l(\frac{k}{\Hbar}\r)^{2(\beta+2)}, 
\eeq
%
where $C$ is a constant of order unity. 
Unlike the massless case, the exact solution to the massive modes
${\bar \mu}_{k}$ is not known in power-law inflation. 
However, it can {be} shown that, for $k_{\rm c}\gg \Hbar$, the WKB
solutions are valid for all $(k\eta)$ over a range of values of
$\beta$ and $k$ of our interest\cite{Shanki-Srir:2004}.  On using the
WKB solutions, we obtain the modified power-spectrum (\ref{eq:psm})
{\em at Hubble exit}\/ to be
\beq
\label{eq:psmfr}
\left[k^3\; {\mathcal P}_{\Psi}(k)\right]_{\rm M}
\simeq  C\,\l(\frac{\Hbar^2}{2\pi^2}\r)\,
\l(\frac{k}{\Hbar}\r)^{2 (\beta + 2)}
\,\l[1 - {\bar C}\,\l(\frac{\Hbar}{k_{c}}\r)\, 
\l(\frac{k}{\Hbar}\r)^{(\beta + 2)}\r], 
\eeq
where ${\bar C}$ is another constant order unity.

The following points are note-worthy regarding the above result:\\
%
\noindent~(i)~Fig.~(1a) contains the plots of the modified and the 
standard spectrum. 
It is evident from the figure that the modified spectrum exhibits 
a suppression of power at the large length scales while it remains 
scale invariant at the small length scales.\\
\noindent~(ii)~Naively, one would expect that TP effects will leave 
their imprints only at the ultra-violet end of the spectrum.  
However, we find that TP effects lead to a modification of the spectrum 
at the infra-red end.  
This can be attributed to the fact that the longer wavelength modes 
leave the Hubble radius at earlier epochs thereby carrying the 
signatures of the TP effects.\\ 
\noindent (iii)~The modified spectrum~(\ref{eq:psmfr}) has some 
similarities to the power spectrum that has been obtained recently 
in non-commutative inflation\cite{Tsujikawa-Maar:2003}.\\
\noindent (iv)~Though the modified spectrum we have obtained exhibits a 
suppression of power around the expected values of $k$, the extent of the 
suppression is far less than that is required to fit the CMB observations. 
In order to illustrate this feature, in Fig. (1b), we have plotted the 
relative power spectrum (i.e. the ratio of the modified spectrum to the 
standard spectrum) of our model and the fit to the WMAP data proposed by 
Contaldi et al.\footnote{Contaldi et al.\cite{Contaldi-Pelo:2003} proposed 
the following form for the primordial spectrum:\\
$~~~~\left[k^3\; {\mathcal P}_{\Psi}(k)\right]_{\rm CPKL}
= A_{\rm s}\, k^{(n_{\rm s}-1)}\,
\biggl[1 - \exp-\l(k/k_\ast\r)^\gamma\biggr]$, where $A_{\rm s}$ and 
$n_{\rm s}$ are the amplitude and index of the standard spectrum,
$k_{\ast}\simeq 5\times 10^{-4}\, {\rm Mpc}^{-1}$ and $\gamma \simeq 3.35$}.
\begin{figure}[h]
$\begin{array}{c@{\hspace{0.2in}}c}
\multicolumn{1}{l}{\mbox{}} &
	\multicolumn{1}{l}{\mbox{}} \\ [-0.53cm]
\epsfxsize=2.45in
\epsffile{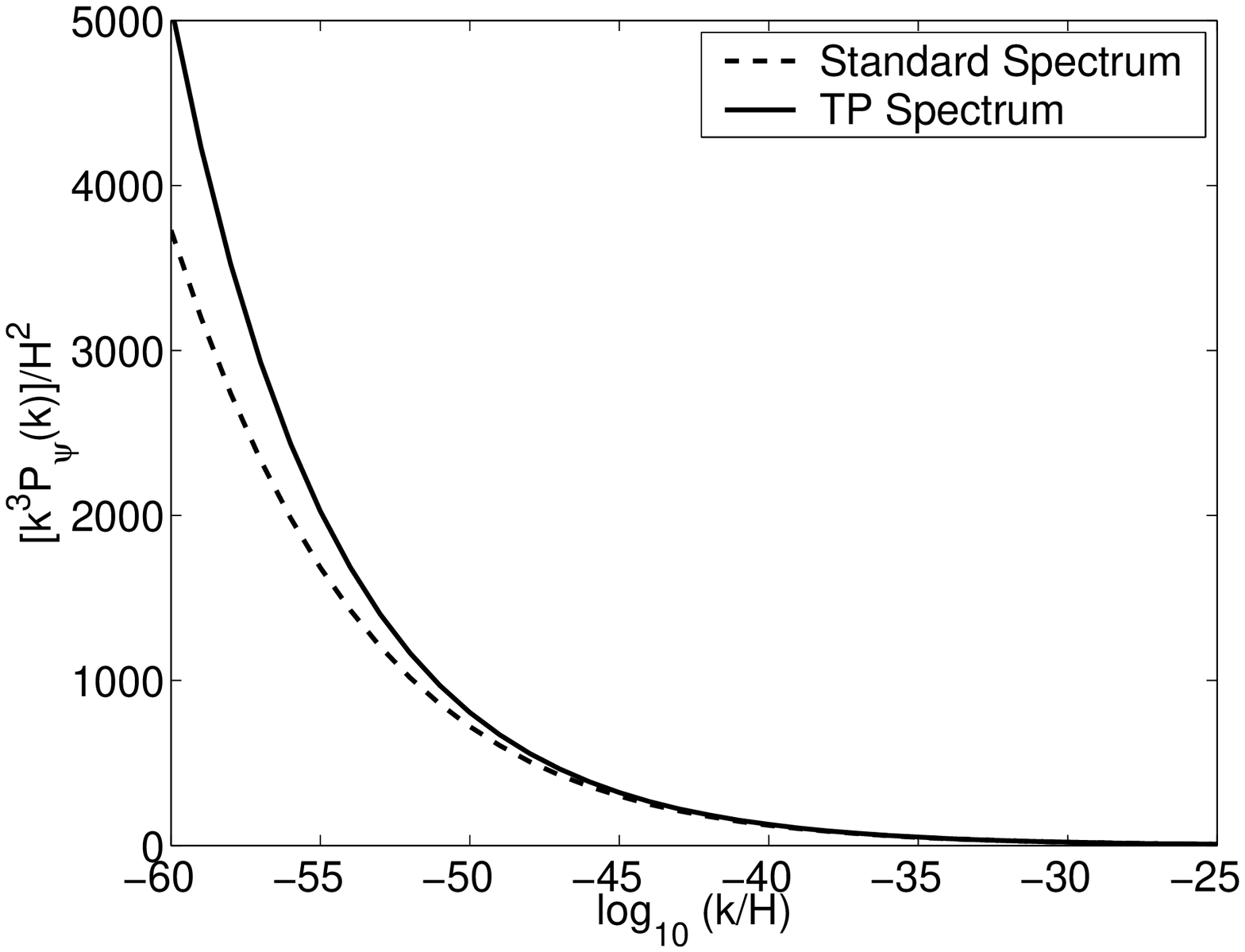} &
	\epsfxsize=2.45in
	\epsffile{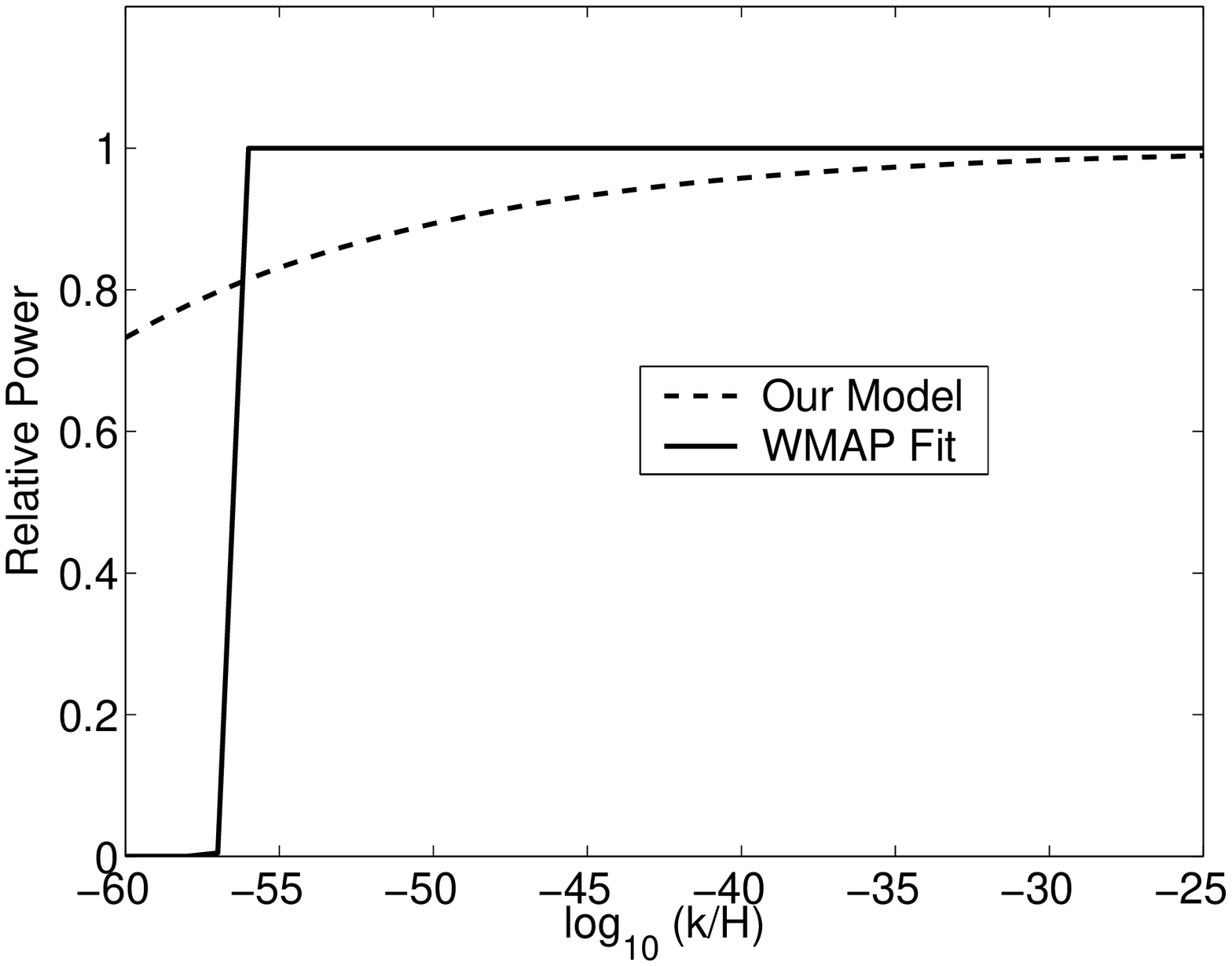} \\ 
\mbox{\bf (a)} & \mbox{\bf (b)}
\end{array}$
\caption{(a)~Plots of the standard power spectrum (solid curve) and the
modified power spectrum (dashed curve), normalized to $\Hbar^2$ for
$\beta = - 2.04$.  (b)~Plots of the relative power of our model
(dashed curve) and Contaldi et al.'s spectrum (solid curve). In both
these plots, the dashed curve corresponds to $(\Hbar/k_{\rm c})= 
10^{-3}$---value chosen so that the WKB approximation is valid around
$k\sim 10^{-4}\, {\rm Mpc}^{-1}$---and $\Hbar=10^{14}\, {\rm GeV}=10^{52}\, 
{\rm Mpc}^{-1}$.}\label{figtest-fig}
\end{figure}

\section{Discussion}

In this proceeding, we have studied the TP effects on the spectrum 
of primordial perturbations in power-law inflation using an approach 
that preserves local Lorentz invariance. 
We assumed that the TP effects modify the standard propagator in a 
particular manner.  
We find that the resulting modified spectrum remains scale invariant 
at the ultra-violet end, but, interestingly, it exhibits a suppression 
of power at the infra-red end---a feature that seems to be necessary
to account for the deficit of power in the lower multipoles of the 
CMB\cite{Peiris-WMAP:2003,Contaldi-Pelo:2003}. 
However, the amount of suppression predicted by our model in power-law 
inflation turns out to be {\it far less}\/ than that seems to be 
required to fit the WMAP data.  
Nevertheless, the loss of power at large scales suggests that the power 
spectrum we have obtained may fit the WMAP data better than the standard 
$\Lambda$CDM model. 
It will be interesting to analyze the implications of our model for WMAP 
data in the context of slow-roll inflation.

\end{document}